\documentclass[11pt]{article}

\usepackage[preprint]{acl}

\usepackage{times}
\usepackage{latexsym}

\usepackage[T1]{fontenc}

\usepackage[utf8]{inputenc}

\usepackage{microtype}
\usepackage{multirow}

\usepackage{inconsolata}

\usepackage{graphicx}

\usepackage{tipa}
\usepackage{float}
\usepackage{booktabs}
\usepackage{makecell} 
\usepackage{threeparttable}
\usepackage{amsmath}

\title{Beyond Unified Models: A Service-Oriented Approach to Low Latency, Context Aware Phonemization for Real Time TTS}

\author{
 \textbf{Mahta Fetrat},
 \textbf{Donya Navabi},
 \textbf{Zahra Dehghanian},
 \textbf{Morteza Abolghasemi},
\\
 \textbf{Hamid R. Rabiee}
\\
\\
 Sharif University of Technology / Tehran, Iran
\\
 \small{
\textbf{Correspondence:}   \texttt{rabiee@sharif.edu}
 }
}

\begin{document}
\maketitle
\begin{abstract}

Lightweight, real-time text-to-speech systems are crucial for accessibility. However, the most efficient TTS models often rely on lightweight phonemizers that struggle with context-dependent challenges. In contrast, more advanced phonemizers with a deeper linguistic understanding typically incur high computational costs, which prevents real-time performance.

This paper examines the trade-off between phonemization quality and inference speed in G2P-aided TTS systems, introducing a practical framework to bridge this gap. We propose lightweight strategies for context-aware phonemization 
and a service-oriented TTS architecture that executes these modules as independent services. This design decouples heavy context-aware components from the core TTS engine, effectively breaking the latency barrier and enabling real-time use of high-quality phonemization models. Experimental results confirm that the proposed system improves pronunciation soundness and linguistic accuracy while maintaining real-time responsiveness, making it well-suited for offline and end-device TTS applications.

\end{abstract}

\section{Introduction}

Text-to-speech (TTS) conversion is a long-established and well-developed task, with a wide range of approaches and architectures proposed over the years. The choice or design of a particular TTS method today depends largely on the specific needs and requirements of the application.

One essential use case for TTS is in screen readers, where the system must operate in real-time, offline, on low-end hardware devices. Users in this setting are exposed to the synthesized voice for long periods every day, so the output must not sound robotic or unpleasant. This scenario imposes three main requirements on the TTS engine: 1) Lightweightness, 2) Real-time performance, and 3) Naturalness.



Unfortunately, there is a clear trade-off among these requirements. Larger and more complex neural models often produce highly natural, human-like speech but require significantly more computational resources and introduce higher inference latency. Conversely, smaller neural models, or traditional rule-based, non-neural systems, are much faster and lighter but lack the capacity to model smooth, natural-sounding human speech.

Simply reducing model size to meet speed and lightweight requirements often degrades speech naturalness. Many recent systems, however, maintain acceptable naturalness by decoupling grapheme-to-phoneme (G2P) conversion from phoneme-to-speech (P2S) synthesis \citep{pipergit, mehta2024matcha, li2025styletts}. Instead of learning an end-to-end text-to-speech mapping, these systems first convert text to phonemes using a lightweight G2P module, then generate speech from the phoneme sequence with a neural synthesizer. This allows the neural component to focus on a narrower task, enabling smaller models and faster inference while preserving reasonable quality.

However, this decoupling makes the overall naturalness and intelligibility of the output heavily dependent on the performance of the G2P module, a task that remains highly challenging for languages with complex or ambiguous phonemization rules.

For example, in Persian, many cases require context-aware phonemization. Two major challenges are:
\begin{enumerate}
    \item Homographs, i.e., words with multiple valid pronunciations depending on context (e.g., the English word \textit{read}, pronounced either \textipa{/ri:d/} or \textipa{/rEd/} depending on tense), and
    \item The Ezafe phoneme, a connecting /e/ sound that appears between grammatically or semantically related words, again determined by context.
\end{enumerate}

\begin{figure}[t]
  \centering
  \includegraphics[width=\columnwidth]{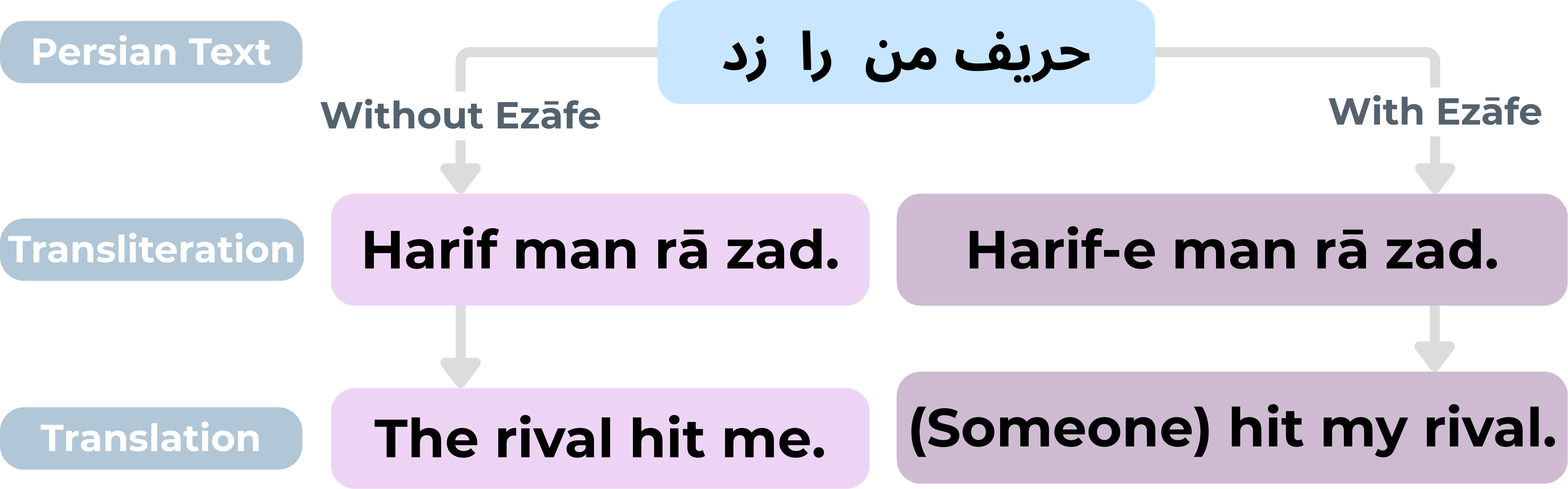}
  \caption{An example of how the Persian Ezafe phoneme (/e/) can change the meaning of a sentence.}
  \label{fig:ezafe-example}
\end{figure}

Figure~\ref{fig:ezafe-example} illustrates how the presence or absence of a single Ezafe phoneme can alter the meaning of a sentence, highlighting the importance of correctly determining it based on context.

Highly non-phonetic and ambiguous languages pose a challenge for lightweight, real-time TTS systems. While embedding a strong, context-aware G2P model could greatly improve pronunciation soundness and correctness, such models are typically large neural networks, and integrating them directly would compromise speed and efficiency. Existing lightweight TTS architectures decouple G2P from P2S, but their G2P modules remain limited for ambiguous languages. Enhancing these modules with context-aware neural models introduces the very latency and computational overhead that lightweight TTS aims to avoid. this is the central challenge addressed in this paper.

In this paper, we propose a method to overcome the latency barrier for incorporating context-aware phonemizers into real-time TTS systems. Our approach combines two complementary strategies: lightweight, statistically driven modules that provide partial context-awareness, and a service-oriented architecture that allows heavier neural phonemizers to run independently, without embedding them directly in the TTS runtime. The core idea is to move beyond the traditional unified TTS design by treating utility modules as independent services, which the main TTS engine can query as needed, avoiding their computational and loading overhead.

Key contributions of this work are as follows:
\begin{enumerate}
    \item Proposing a service-oriented approach for integrating neural components into real-time TTS systems,
    \item Presenting a service-oriented adaptation of the well-known PiperTTS architecture,
    \item Introducing a lightweight, fast, and context-aware phonemizer tailored to Persian phonemization challenges, an enhanced version of the existing eSpeak phonemizer, and
    \item Providing a new Persian voice for Piper, trained on the largest publicly available Persian TTS dataset to date.
\end{enumerate}

\section{Related Works}

\subsection{TTS approaches}

TTS is a longstanding task that has been in existence since 1939~\cite{dudley1939synthetic}. It began with rule-based methods that utilized hand-written pronunciation and prosody rules, along with simple formant/articulatory synthesis, to generate speech~\cite{klatt1980software,klatt1987review}. Then it proceeded to the next generation, utilizing concatenative unit-selection~\cite{sagisaka1988speech,hunt1996unit,black1997automatically} and later statistical parametric systems (e.g., HMM-based acoustic models with vocoders)~\cite{tokuda2000speech,zen2009statistical}, which improved stability and footprint but still had limitations in terms of naturalness. Like many other tasks, it then evolved into deep-learning-based methods like sequence-to-sequence acoustic models (Tacotron-style)~\cite{wang2017tacotron,shen2018natural} paired with neural vocoders (WaveNet/flow/GAN)~\cite{van2016wavenet,prenger2019waveglow,yamamoto2020parallel} and fully end-to-end models such as VITS~\cite{kim2021conditional} and non-autoregressive FastSpeech-style models~\cite{ren2019fastspeech,ren2020fastspeech}; these can be grouped by architecture families (autoregressive, non-autoregressive, flow-based, diffusion-based)~\cite{kim2020glow,popov2021grad,kim2022guided,mehta2024matcha}. And most recently, it is performed by large language models, such as commercial or research foundation TTS systems (e.g., VALL-E/Bark-style and TTS components integrated into general LLM stacks)~\cite{wang2023neural,le2023voicebox}, which offer strong quality but usually require online GPU inference.

In this paper, our focus is on TTS architectures suitable for offline, real-time, end-device applications. Therefore, we limit our discussion to models that can operate efficiently in CPU-first, low-latency settings.

\begin{figure}[t]
  \includegraphics[width=\columnwidth]{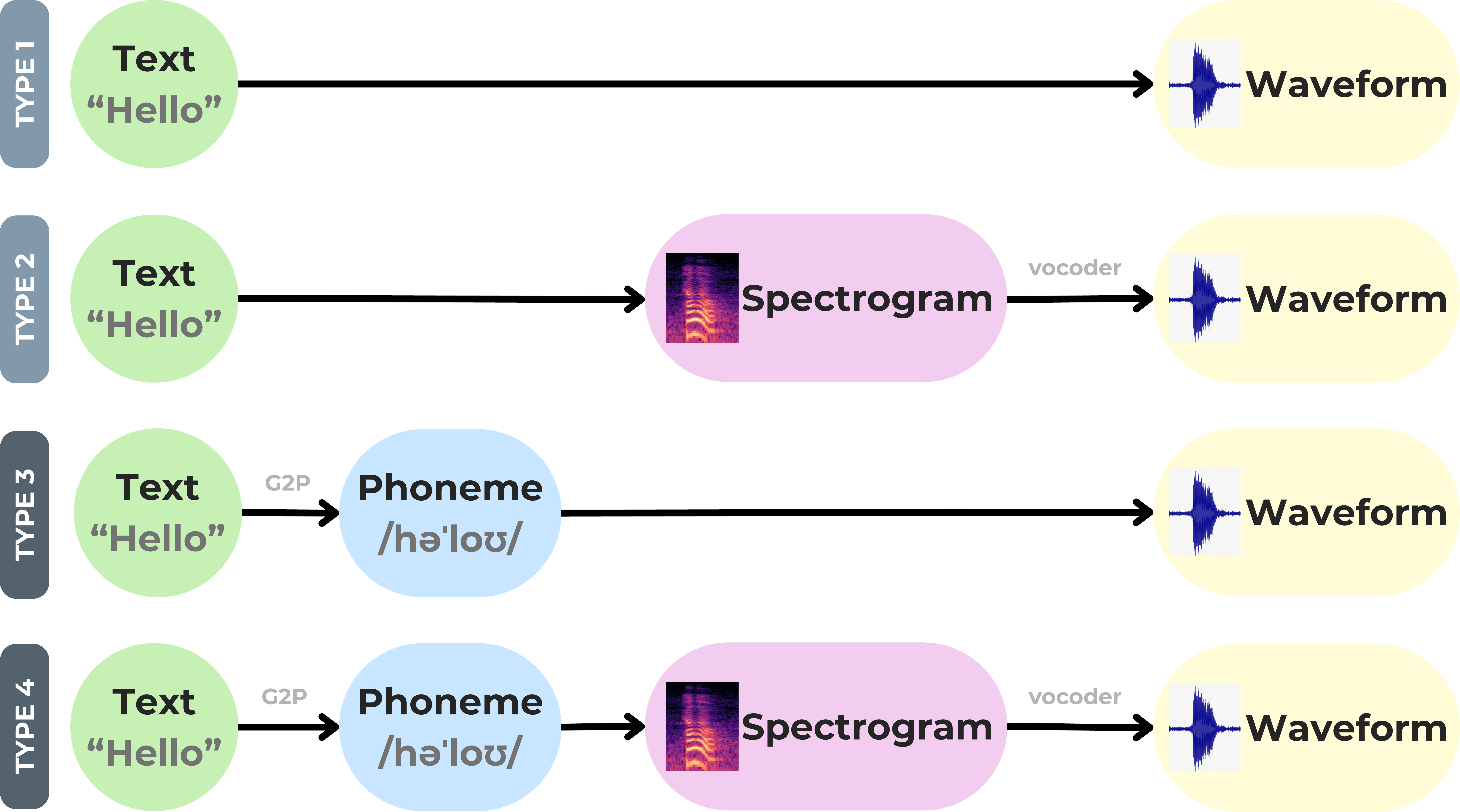}
  \caption{Four common granularity levels in TTS architectures, differing by intermediate representations.}   
  \label{fig:task-granularity}
\end{figure}

In practice, we can narrow our focus to TTS architectures that include a distinct phonemization stage. Broadly, modern TTS systems can be categorized into four levels of architectural granularity (Figure~\ref{fig:task-granularity}). At one extreme, fully end-to-end models map raw text directly to waveform; at the other, highly modular pipelines decompose the task into explicit stages: first converting text to phonemes, then generating a spectrogram, and finally synthesizing the waveform through a vocoder. Understanding the degree of granularity in a given TTS architecture provides insight into the complexity of the problem the model must solve, the capacity it may require, and the latency implications of its intermediate components. This perspective is essential when selecting an appropriate model for applications with constraints such as low-latency or limited compute. 

VITS, FastSpeech-family models, Glow-TTS, and other flow-based systems, as well as recent diffusion/consistency approaches (e.g., Matcha-like designs), are the most relevant to our goals~\cite{kim2021conditional,ren2019fastspeech,ren2020fastspeech,kim2020glow,mehta2024matcha,Piper:2023}.In summary: VITS merges acoustic modeling and vocoding and offers good quality with low-latency; FastSpeech generates spectrograms in parallel and is very fast with a light vocoder; flow-based models enable stable alignments and parallel inference; diffusion/consistency models improve robustness and quality with careful inference schedules~\cite{kim2021conditional,ren2019fastspeech,kim2020glow,mehta2024matcha,li2023styletts}.

Piper architecture, which is the baseline model in this study, is closely related to these families and is based on VITS with practical improvements for deployment~\cite{ZachB100:2023PiperGuide}. It uses a modular structure with a G2P front-end and a phoneme-to-speech (P2S) neural back-end. By moving the G2P step outside the neural model (typically using a lightweight rule-based phonemizer such as eSpeak-ng~\cite{eSpeakNG:2024}) and exporting models to ONNX for CPU-friendly inference, Piper reduces model size, and improves speed while maintaining acceptable naturalness. For a more detailed justification for choosing Piper as our baseline, please refer to Appendix~\ref{appendix:piper}.

\subsection{G2P Tools}

Since a central focus of this study is enhancing the G2P component of a TTS system, we briefly review the relevant literature and available tools.

G2P methods have evolved in parallel with TTS systems. Early rule-based approaches and pronunciation lexicons were compact and predictable but struggled with out-of-vocabulary words and context-dependent pronunciations. Statistical methods such as finite-state and n-gram letter-to-sound models and CRF-based taggers generalized better while remaining relatively lightweight~\cite{beesley2003finite,bisani2008joint,jiampojamarn2007applying}. Neural models, including RNNs and Transformers, have since achieved state-of-the-art accuracy by capturing longer-range dependencies~\cite{yao2015sequence,vaswani2017attention}, but they typically require more compute and memory than lightweight statistical or rule-based methods~\cite{G2pE:2019}.

In the case of Persian, several non-scholarly G2P implementations exist on platforms such as GitHub~\citep{persian-phonemizer, PasaOpasen, azamrabiee, mohammadhasan, mortensen2018epitran, sajadalipour7}. A recent benchmark study evaluated these tools and found their performance to be unsatisfactory, reporting phoneme error rates (PER) between 15-50\%, homograph disambiguation accuracy below random baseline, and Ezafe detection F1 scores ranging from 6-60\%~\citep{qharabagh2025llm}. Subsequently, a Persian LLM-powered G2P model was introduced that substantially improved these metrics~\citep{qharabagh2025llm}. Nevertheless, such models are not suitable for free, offline, or real-time use, the key constraints of our target applications.

Building on those findings, another study leveraged the outputs of the LLM-based system to create a new dataset and train two open-source, offline G2P models~\citep{qharabagh2025fast}: Homo-GE2PE~\citep{HomoGE2PEPersian} and HomoFast eSpeak~\citep{HomoFast-eSpeak}. Homo-GE2PE is a high-quality neural G2P model that performs well across PER, homograph disambiguation, and Ezafe detection. HomoFast eSpeak, in contrast, is entirely non-neural and extremely fast, achieving good PER and homograph accuracy but offering limited Ezafe detection capability due to its lack of linguistic modeling.

\subsection{Decoupled TTSs}

To the best of our knowledge, no prior work has proposed structuring a TTS system so that some of its internal submodules operate as independent services to take advantage of modular decoupling. While several studies and open-source projects provide complete TTS systems as API-based services~\cite{Festival:ServerAPI, marytts, Pipecat:PiperHTTP, LlamaEdge:PiperAPIServer}, this should not be confused with the approach presented here.

Our work differs fundamentally from these systems: rather than exposing the entire synthesizer as a remote service, we implement a service-based decomposition within the TTS pipeline itself. This design decouples computationally heavy, higher-latency modules, such as context-aware phonemization components, from the lightweight inference core, improving overall responsiveness and enabling real-time performance.

\section{Methodology}

As discussed earlier, our baseline system is Piper, which adopts a two-stage pipeline (Type~3 granularity in Figure~\ref{fig:task-granularity}):
(1) a text-to-phoneme conversion step implemented with the eSpeak phonemizer, followed by
(2) a neural phoneme-to-speech (P2S) model that synthesizes the waveform without a separate vocoder.
The primary focus of this work is to strengthen the first stage by introducing context-aware phonemization and to address the practical challenges that arise when integrating this improved G2P component into the complete TTS pipeline.

\begin{figure*}[t]
    \centering
    \includegraphics[width=0.8\textwidth]{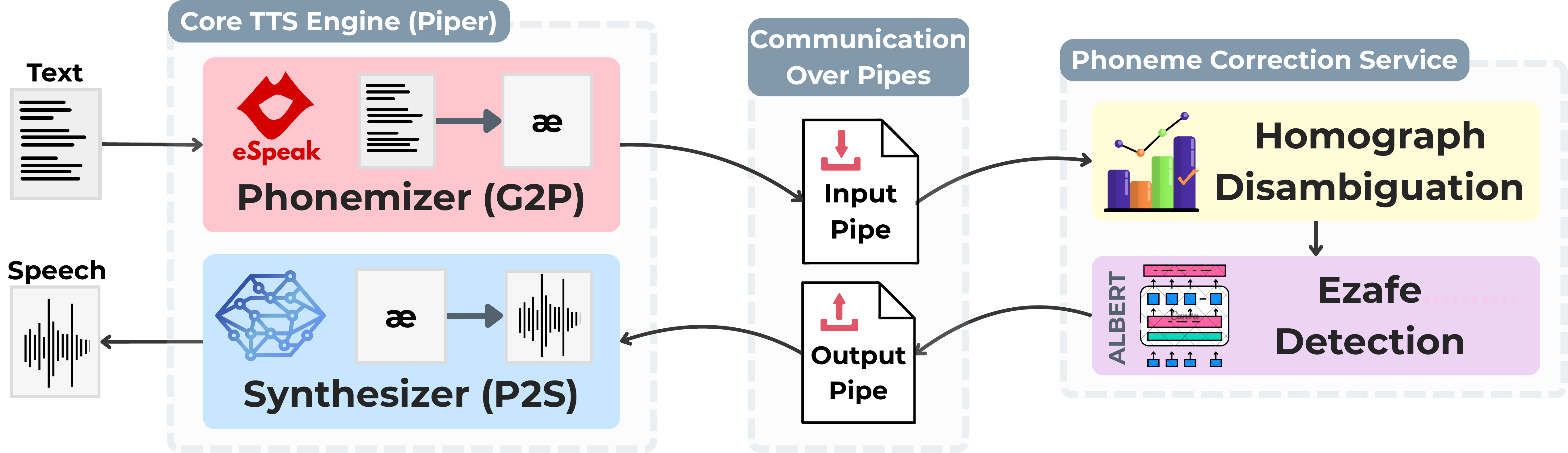}
    \caption{The proposed service-based architecture for context-aware TTS.}
    \label{fig:integration}
\end{figure*}

The default phonemizer in PiperTTS, eSpeak, is a rule-based system relying on dictionary lookups and hardcoded linguistic rules. This design introduces weaknesses for languages that require context-aware phonemization, particularly in handling homograph disambiguation and Ezafe detection in Persian. We propose two complementary families of solutions to address these challenges.

\subsection{Statistical Context-Awareness}

Context-awareness can be introduced in lightweight TTS systems using simple statistical methods. Certain phonemization tasks, such as homograph disambiguation, can be addressed with shallow contextual statistics instead of heavy neural models.

\citet{qharabagh2025fast} showed that a method based on word co-occurrence distributions can improve homograph disambiguation accuracy by up to 30 percentage points. Their approach constructs a database of homographs and their commonly associated context words, selecting the pronunciation with the highest contextual overlap for a given input. We adopt this lightweight strategy to enhance PiperTTS’s phonemizer without adding computational overhead or latency.

\subsection{Distilled Linguistic Knowledge}

Certain aspects of phonemization require deeper linguistic understanding, such as detecting the Ezafe phoneme, which depends on grammatical and semantic relations between words. However, full-scale language understanding is not necessary for this task. Task-specific, lightweight neural models can be effectively trained via knowledge distillation from larger models.

In our case, Ezafe detection can be viewed as a subtask of part-of-speech (POS) tagging. The SpaCy POS tagger for Persian \citep{spacy_roshan} is reported to achieve an F1 score of 0.99298 on Ezafe tagging but is relatively heavy and slow during inference (Table~\ref{tab:ezafe-eval}). To obtain a lighter alternative, we distilled the Ezafe tagging knowledge of the SpaCy model into a smaller model based on ALBERT \citep{lan2019albert}.

We created a labeled dataset from the text portion of the ManaTTS corpus \citep{qharabagh2025manatts} by automatically annotating Ezafe tags using the SpaCy tagger’s predictions. A pretrained Persian ALBERT model \citep{albertfa_zwnj_base_v2} was then fine-tuned on this data, producing a smaller, faster model with performance nearly comparable to the original tagger (Table~\ref{tab:ezafe-eval}). For efficient CPU-based inference and reduced memory usage, the distilled model was exported to ONNX.

\subsection{Service-Based Integration}

When all components of a TTS system are integrated into a single unified runtime, the individual loading and inference delays of each module accumulate, resulting in a significant overall latency. To overcome this bottleneck, we adopted a service-oriented architecture, setting up utility modules as independent, persistent services running in separate processes. The core TTS module communicates with these services using inter-process communication (IPC) via piped input and output files. This design decouples the initialization of independent modules and significantly reduces latency during inference.

In our setup, the context-aware phonemization components operate as a dedicated service, with the core TTS engine interacting through two file pipes (input and output). For each input text, the core TTS module first generates an initial phoneme sequence using its default phonemizer (PiperTTS’s eSpeak-based component). This sequence is then sent to the context-aware phonemization service, where it undergoes two refinement stages: the homograph disambiguation module corrects potential mispronunciations, and the Ezafe detection model inserts any missing Ezafe phonemes. The enhanced phoneme sequence is then returned to the core TTS engine and passed to the phoneme-to-speech model, which synthesizes the final audio output. Figure~\ref{fig:integration} illustrates this service-based setup for the context-aware TTS proposed in this study.

Finally, we fine-tuned the P2S model on the phoneme sequences produced by the enhanced phonemizer. Fine-tuning was carried out for 1,000 epochs with a batch size of 32 on a workstation equipped with an NVIDIA A100-SXM4 GPU (80 GB) and an Intel Xeon Platinum 8380 CPU with 1 TB of system memory, using the Persian ManaTTS dataset \citep{qharabagh2025manatts}. This step was crucial for enabling the model to correctly handle Ezafe phonemes and to distinguish between homographs that differ by only a few phonemes, rather than biasing toward the most frequent pronunciations observed in baseline models.

\begin{table*}[t]
\centering
\renewcommand{\arraystretch}{1.12}
\setlength{\tabcolsep}{6pt}
\begin{tabular}{lcccccc}
\toprule
\textbf{Model} & \textbf{PER} & \textbf{Ezafe F1} & \textbf{Homograph} & \multicolumn{2}{c}{\textbf{RTF $\downarrow$}} \\
 & (\% $\downarrow$) & (\% $\uparrow$) & Acc. (\% $\uparrow$) & Direct Call & Service-Based \\
\midrule
\makecell[l]{MatchaTTS \\ \citep{persian_matcha_tts}} &
  6.32 $\pm$ 0.00 &
  19.58 $\pm$ 0.00 &
  43.87 $\pm$ 0.00 &
  0.185 $\pm$ 0.051 &
  -- \\
\makecell[l]{GlowTTS \\ \citep{Kamtera_glow_tts}} &
  6.61 $\pm$ 0.00 &
  19.96 $\pm$ 0.00 &
  43.87 $\pm$ 0.00 &
  1.364 $\pm$ 0.705 &
  -- \\
\makecell[l]{Piper (Base) \\ \citep{persian_piper}} &
  6.32 $\pm$ 0.00 &
  19.58 $\pm$ 0.00 &
  43.87 $\pm$ 0.00 &
  \textbf{0.153 $\pm$ 0.012} &
  -- \\
\midrule
\makecell[l]{Piper + Neural G2P} &
  \underline{4.95 $\pm$ 0.68} &
  \underline{87.70 $\pm$ 0.78} &
  \underline{74.53 $\pm$ 0.39} &
  3.840 $\pm$ 0.415 &
  0.396 $\pm$ 0.095 \\
\makecell[l]{\textbf{Piper + LCA G2P}} &
  \textbf{4.80 $\pm$ 1.06} &
  \textbf{90.08 $\pm$ 0.72} &
  \textbf{77.67 $\pm$ 0.22} &
  5.519 $\pm$ 0.984 &
  \underline{0.167 $\pm$ 0.015} \\
\bottomrule
\end{tabular}
\caption{Comparison of phonemization accuracy and inference speed across baseline and proposed TTS models.}
\label{tab:results}
\end{table*}

\begin{table*}[t]
\centering
\begin{tabular}{lcccccc}
\toprule
\textbf{Model} & \textbf{Params} & \textbf{Memory} & \textbf{Disk} & \textbf{Ezafe F1} & \textbf{Avg. Inf. Time} \\
 & (Millions $\downarrow$)  & (MB $\downarrow$) & (MB $\downarrow$)  & (\% $\uparrow$) & (s $\downarrow$) \\
\midrule
SpaCy \citep{spacy_roshan} &
162.84 & 621.19 & 1258.49 &
\textbf{97.67 $\pm$ 0.00} &
0.110 $\pm$ 0.004 \\

ALBERT-based (Ours) &
\textbf{11.09} & \textbf{42.32} & \textbf{41.38} &
94.19 $\pm$ 0.00 &
\textbf{0.037 $\pm$ 0.001} \\
\bottomrule
\end{tabular}
\caption{Comparison between the SpaCy teacher model and the distilled ALBERT-based Ezafe detector.}
\label{tab:ezafe-eval}
\end{table*}

\section{Experiments}

In this section, we evaluate how the proposed context-aware phonemization modules improve a rule-based phonemizer’s accuracy and affect inference speed. While context-aware modules enhance phonemization quality, they also introduce additional computational load that can increase latency. We therefore assess their performance within our service-based framework, which decouples these heavier components and restores real-time operation without compromising phonemization improvements.

Our enhanced system, Piper equipped with the proposed lightweight context-aware phonemization components, is referred to as "Piper + LCA-G2P", where LCA stands for Lightweight Context-Aware. To demonstrate the framework’s ability to handle heavier models, we also integrated the state-of-the-art Persian G2P model, Homo-GE2PE \citep{qharabagh2025fast}, which handles both homograph disambiguation and Ezafe detection. This setup, "Piper + Neural G2P", uses a substantially larger model (300M parameters) compared to Piper’s lightweight 15-20M parameters, showing that the framework can accommodate computationally intensive neural components while maintaining real-time performance.

All experiments evaluating real-time factor (RTF)\footnote{RTF, or real-time factor, is the ratio of audio synthesis time to the duration of the generated audio. For instance, an RTF of 0.2 indicates synthesis five times faster than real-time.} were conducted on a typical end-device configuration: a Windows system with a 12th Gen Intel Core i7-1255U CPU (10 cores, 1.7 GHz) and 16 GB of RAM, running CPU-only inference. This setup demonstrates the system’s suitability for offline, low-latency, and real-time applications, without relying on GPU acceleration. The results are summarized in Table~\ref{tab:results}.

\subsection{Mean Opinion Score}

The enhanced soundness and context-awareness of the phonemization process, along with the subsequent fine-tuning of the phoneme-to-speech (P2S) engine, are expected to improve the overall naturalness of the generated speech. Table~\ref{tab:mos} presents the Mean Opinion Score (MOS) results for the baseline and enhanced TTS systems, as well as the reference natural speech.

Table~\ref{tab:mos} shows the average Mean Opinion Score (MOS) for the baseline and enhanced TTS systems, alongside natural speech, based on evaluations from 16 native Persian speakers across seven utterances. For full details of the experiment, please refer to Appendix~\ref{appendix:mos-details}.

\begin{table}[t]
\centering
\renewcommand{\arraystretch}{1.15}
\setlength{\tabcolsep}{10pt}
\begin{tabular}{l c}
\toprule
\textbf{Source} & \textbf{MOS $\uparrow$} \\
\midrule
\textbf{Glow} \citep{Kamtera_glow_tts} & 1.30 $\pm$ 0.75 \\
\textbf{Matcha} \citep{persian_matcha_tts} & 2.54 $\pm$ 0.99 \\
\textbf{Piper (Base)} \citep{persian_piper} & 2.41 $\pm$ 0.84 \\
\textbf{Piper + LCA G2P (Ours)} & \textbf{3.14 $\pm$ 1.00} \\
\textbf{Natural Speech} & 4.21 $\pm$ 0.97 \\
\bottomrule
\end{tabular}
\caption{MOS of the baseline and enhanced TTS system compared to natural speech.}
\label{tab:mos}
\end{table}

\subsection{Ezafe Detection Module Evaluation}

This section presents the experiments conducted on the distilled Ezafe detection module, demonstrating that it achieves a substantial reduction in size and computational overhead while retaining the strong performance of its teacher model. All experiments were conducted on a CPU environment in Google Colab. The results are summarized in Table~\ref{tab:ezafe-eval}.

\section{Conclusion}

This study addressed the fundamental trade-off between speed, lightweightness, and context-aware phonemization in G2P-aided TTS systems. We proposed practical approaches to mitigate this challenge, including methods for developing auxiliary context-aware modules that are inherently lighter and faster, as well as introducing a service-based architecture that enables their efficient integration into real-time TTS pipelines.

The proposed framework demonstrated that it is possible to achieve enhanced phonemization accuracy without compromising real-time performance. By decoupling heavy context-aware components from the core runtime and executing them as independent services, the system maintained low-latency while significantly improving the overall soundness of the synthesized speech. These characteristics make the architecture particularly suitable for offline, end-device, and low-latency applications such as screen readers.

All source code, models, and experimental results from this work are publicly available.
\footnote{\url{https://github.com/MahtaFetrat/Piper-with-LCA-Phonemizer}}

\section*{Limitations}

Even with fully corrected phoneme sequences in the TTS system, achieving complete naturalness remains out of reach. This is primarily because lightweight TTS models have limited capacity in the phoneme-to-speech component, which is typically insufficient to fully capture or reproduce higher-level prosodic and expressive features. As a result, the overall perceived naturalness cannot reach its maximum potential. Further research is needed to improve these aspects of naturalness while maintaining the desired properties of speed and lightweight design.

Another consideration is that, from a perceptual standpoint, naturalness is more closely associated with qualities such as smoothness, noiselessness, and accurate intonation and stress patterns. Correct phonemization primarily affects pronunciation soundness and only indirectly contributes to perceived naturalness. It may therefore be valuable to design subjective evaluation protocols that separate the assessment of phonemization accuracy from other dimensions of naturalness, such as prosody and fluency.

Another limitation, or rather an avenue for future enhancement, lies in the service-based setup itself. Now that several components are decoupled from the core TTS engine, additional optimization strategies can be applied to the service layer. For example, implementing request-level parallelism or asynchronous processing could further reduce overall system latency and improve scalability.

\bibliography{custom}

\appendix

\section{Selection Criteria for PiperTTS}
\label{appendix:piper}

Considering the requirements of the TTS system in this study and the models reviewed in the related work section, we found PiperTTS to be the most suitable architecture for our needs. PiperTTS has been available for several years and has gained significant community attention and contributions \citep{pipergit}. It exhibits several characteristics that align with the requirements and objectives of this research:

\begin{itemize}
    \item \textbf{Lightweightness:} One of PiperTTS’s core strengths is its ability to run efficiently on CPUs and even low-end devices such as Raspberry Pi. Its ONNX runtime export enables lightweight deployment on various hardware platforms.
    \item \textbf{Speed:} PiperTTS demonstrates high inference speed, achieving a real-time factor (RTF) of approximately~0.2.
    \item \textbf{Naturalness:} The model provides medium to high perceptual quality, as verified through publicly available checkpoints \citep{gyroing}.
    \item \textbf{Accessibility Integration:} PiperTTS has already been integrated into the open-source NVDA screen reader, which embeds TTS engines through add-ons called synthesizer drivers. An established Piper synthesizer driver is publicly available \citep{sonata}.
    \item \textbf{Open Source:} Being open source, PiperTTS facilitates the development of accessible tools and enables contributions to a field that currently lacks substantial open research.
    \item \textbf{Persian Support:} The model has an active Persian-speaking community, with available checkpoints and established Persian training setups.
\end{itemize}

Given these factors, PiperTTS was selected as the base TTS architecture for this work.

\begin{figure*}[t]
\centering
\includegraphics[width=\textwidth]{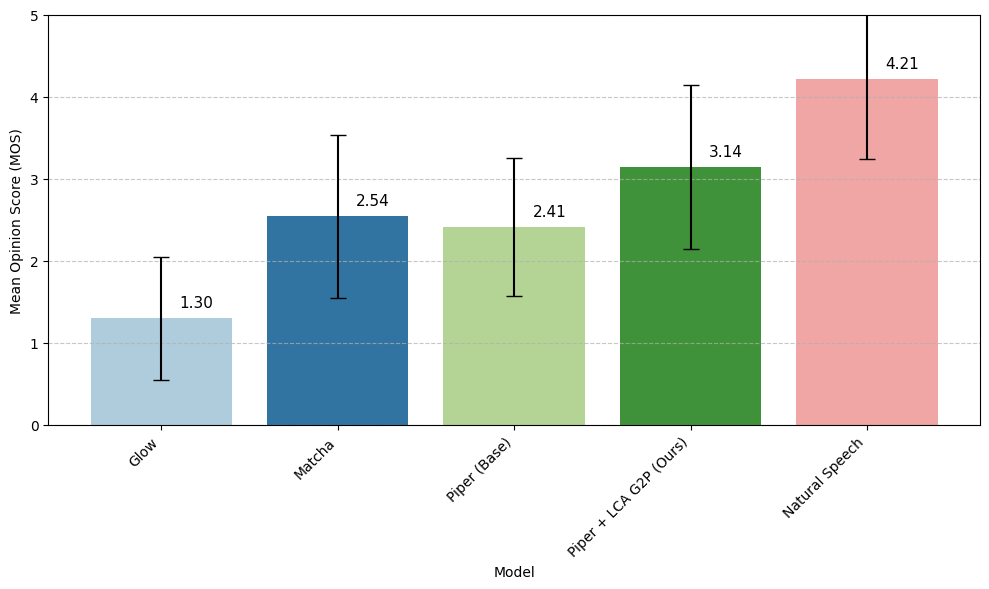}
\caption{Overall average MOS across all seven utterances for each TTS system and natural speech, highlighting the improved naturalness of the enhanced Piper system.}
\label{fig:mos-barplot}
\end{figure*}

\begin{table*}[h!]
\centering
\begin{tabular}{lcccccc}
\toprule
 & & \textbf{Piper + LCA} & \textbf{Natural} & \textbf{Glow} & \textbf{Matcha} & \textbf{Piper (Base)} \\
\cmidrule(lr){2-7}

\multirow{2}{*}{\textbf{Utterance 1}}
  & \textbf{MOS} 
  & 2.94 $\pm$ 0.68 & 4.12 $\pm$ 0.50 & 1.19 $\pm$ 0.54 & 2.25 $\pm$ 0.77 & 2.38 $\pm$ 0.81 \\
  & \textbf{Order} 
  & 5 & 1 & 2 & 3 & 4 \\
\cmidrule(lr){2-7}

\multirow{2}{*}{\textbf{Utterance 2}}
  & \textbf{MOS} 
  & 3.75 $\pm$ 0.93 & 4.25 $\pm$ 1.00 & 2.00 $\pm$ 1.03 & 2.62 $\pm$ 1.09 & 2.38 $\pm$ 0.89 \\
  & \textbf{Order} 
  & 3 & 2 & 4 & 1 & 5 \\
\cmidrule(lr){2-7}

\multirow{2}{*}{\textbf{Utterance 3}}
  & \textbf{MOS} 
  & 3.19 $\pm$ 0.91 & 4.88 $\pm$ 0.34 & 1.12 $\pm$ 0.50 & 2.44 $\pm$ 0.96 & 2.12 $\pm$ 0.62 \\
  & \textbf{Order} 
  & 4 & 3 & 2 & 5 & 1 \\
\cmidrule(lr){2-7}

\multirow{2}{*}{\textbf{Utterance 4}}
  & \textbf{MOS} 
  & 2.81 $\pm$ 0.98 & 4.56 $\pm$ 1.03 & 1.12 $\pm$ 0.34 & 2.50 $\pm$ 1.03 & 2.19 $\pm$ 1.05 \\
  & \textbf{Order} 
  & 2 & 5 & 4 & 3 & 1 \\
\cmidrule(lr){2-7}

\multirow{2}{*}{\textbf{Utterance 5}}
  & \textbf{MOS} 
  & 2.62 $\pm$ 1.15 & 4.31 $\pm$ 0.70 & 1.19 $\pm$ 0.75 & 2.69 $\pm$ 1.08 & 3.00 $\pm$ 0.73 \\
  & \textbf{Order} 
  & 2 & 4 & 1 & 5 & 3 \\
\cmidrule(lr){2-7}

\multirow{2}{*}{\textbf{Utterance 6}}
  & \textbf{MOS} 
  & 3.69 $\pm$ 0.87 & 3.81 $\pm$ 1.22 & 1.25 $\pm$ 0.77 & 2.62 $\pm$ 1.02 & 2.62 $\pm$ 0.81 \\
  & \textbf{Order} 
  & 2 & 1 & 5 & 3 & 4 \\
\cmidrule(lr){2-7}

\multirow{2}{*}{\textbf{Utterance 7}}
  & \textbf{MOS} 
  & 3.00 $\pm$ 1.03 & 3.56 $\pm$ 1.15 & 1.25 $\pm$ 0.77 & 2.62 $\pm$ 1.09 & 2.19 $\pm$ 0.75 \\
  & \textbf{Order} 
  & 1 & 2 & 4 & 5 & 3 \\
\bottomrule
\end{tabular}
\caption{Per-utterance MOS (mean $\pm$ std) for each source. ``Order'' shows the presentation order (1--5) of the sources for that utterance (randomized per utterance).}
\label{tb:utterances}
\end{table*}

\begin{figure*}[t]
  \centering
  \includegraphics[width=0.6\textwidth]{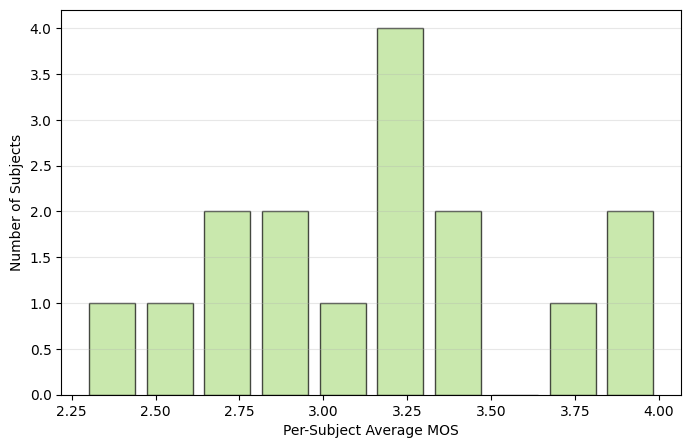}
  \caption{Distribution of MOS ratings assigned by participants to the enhanced TTS system  (Piper + LCA-G2P).}
  \label{fig:mos-distribution}
\end{figure*}

\begin{figure*}[t]
\centering
\includegraphics[width=0.9\textwidth]{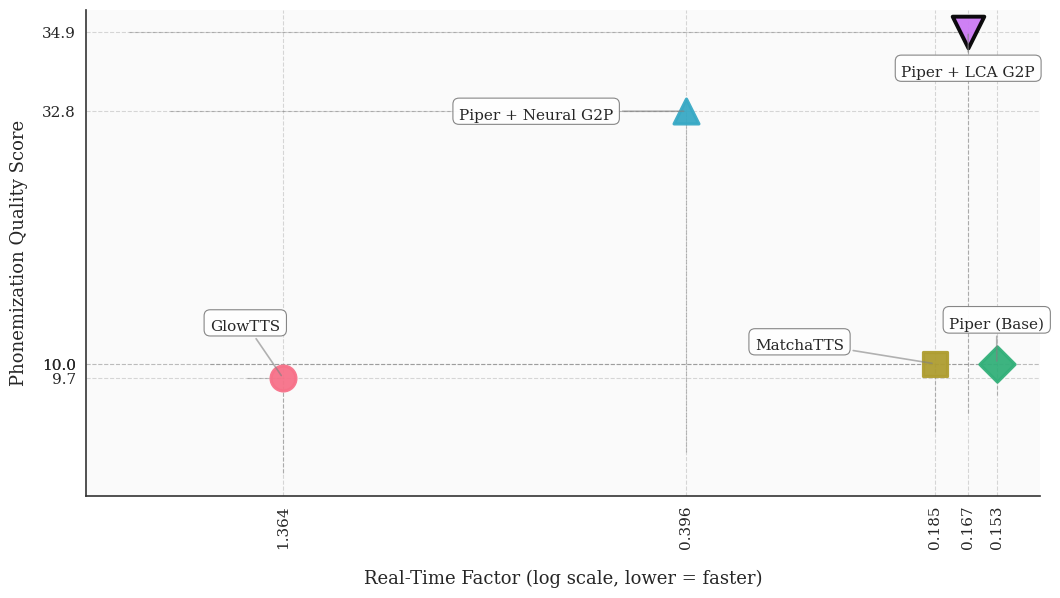}
\caption{Trade-off between inference speed (RTF, log scale) and phonemization quality (composite metric) for various TTS models. The top-right region indicates models that are both fast and high-quality.}
\label{fig:speed-quality}
\end{figure*}

\section{Mean Opinion Score Details}
\label{appendix:mos-details}

To evaluate perceived naturalness, we selected seven utterances from a recent issue of the online monthly magazine \emph{Nasl-e-Mana}, a publication for the blind community and the source of the publicly available ManaTTS dataset. The chosen issue contained content not included in the ManaTTS corpus used for fine-tuning the phoneme-to-speech model.

For each utterance, audio was generated using five sources: two open-source lightweight Persian TTS models (GlowTTS and MatchaTTS), the baseline PiperTTS model, our enhanced Piper system, and the corresponding natural speech recordings. The audio samples for all utterances are provided in the repository's samples directory.
\footnote{\url{https://github.com/MahtaFetrat/Piper-with-LCA-Phonemizer}}
To avoid bias based on overall model reputation or perceived quality, the order of the five sources was independently shuffled for each utterance.

Sixteen native Persian speakers were asked to rate the naturalness of each audio clip on a scale from 1 to 5 (MOS), with 5 indicating the most natural pronunciation. Participants were instructed as follows (translated from Persian):

"Please listen to each audio clip and rate its naturalness on a scale from 1 to 5. A score of 5 corresponds to fully natural pronunciation, while a score of 1 corresponds to the least natural or highly robotic pronunciation. Lower your rating if you detect any unnatural intonation, mispronunciation, or mechanical quality."

The resulting overall MOS values, averaged across all utterances, are shown in Figure~\ref{fig:mos-barplot}. Detailed MOS results per utterance, including the shuffled order of sources, are provided in Table~\ref{tb:utterances}. Standard deviations are reported to reflect inter-subject variability. Additionally, the distribution of MOS scores assigned to our enhanced model by individual participants is illustrated in Figure~\ref{fig:mos-distribution}.

\section*{Additional Figures}

Visualizing experimental findings can provide valuable insight. A central concern of this study is the trade-off between inference speed and phonemization quality: as the quality of TTS improves, especially with the aid of context-aware and linguistically-informed phonemization tools, inference typically becomes slower. Our work proposes a service-based architecture that mitigates this trade-off, allowing models to achieve both reasonable speed and improved phonemization quality.

Figure~\ref{fig:speed-quality} illustrates the performance of the evaluated models along two axes: speed and quality. Inference speed is represented by the real-time factor (RTF), displayed on a logarithmic scale. For quality, we define a composite metric that combines the key context-aware phonemization challenges studied in this work:

\begin{equation}
\text{G2P\ Quality} = \frac{\text{Ezafe\ F1} + \text{Homograph\ Acc.}}{\text{PER}}
\end{equation}

This formulation rewards higher Ezafe F1 and homograph accuracy while penalizing higher phoneme error rate, providing an intuitive measure aligned with our goals. The quality values are schematically arranged for visualization, maintaining relative correctness.

In this plot, models that are both fast and high-quality appear in the top-right corner. As expected, our proposed enhanced version, "Piper + LCA-G2P", occupies this position, demonstrating that it maintains strong inference speed while substantially improving phonemization quality.

\end{document}